# Influence of relative sea-level rise, meteoritic water infiltration and rock weathering on giant volcanic landslides: theory and real cases

Julien Gargani
Université Paris-Saclay, Geops, CNRS, Orsay, France
julien.gargani@universite-paris-saclay.fr

**Abstract**
Recent studies have shown that giant landslides seem to be correlated with climatic variations. Nevertheless, the precise processes that are involved in this phenomenon need to be better constrained. In this study, the causes of giant landslides are investigated using a modeling approach. Here, we show that the effect of meteoritic water infiltration could be discriminated from that of sea-level rise on triggering paleolandslides. It is possible to identify the cause of coastal paleolandslides based on the age of occurrence and comparison with climatic signals when glacial maxima are more humid than during interglacial times, as in Polynesia and East Equatorial Africa, but not in other cases (Caribbean, Indonesia). The role of pore pressure variations and sea-water loading variations has been discussed. The interaction between the relative sea-level rise, preexisting relief and deep weak structure due to the presence of highly weathered lavas may trigger the conditions for a large landslide. Highly weathered lavas have very low friction angles at depth in volcanic islands. When volcanoes are still actives, pressure variation of the magma chamber caused by sea-level lowering is expected to play a significant role in destabilization of the relief. Competing processes in real cases cause difficulties to discriminate between these processes.

## 1. Introduction

Landslides are one of the main processes that destroy relief and displace material. Catastrophic landslides are often expected in areas of active seismicity (Keefer, 1994) or in the context of significant volcanic activity (Carracedo et al., 1999; Blahüt et al., 2019). In the case of active seismicity, the acceleration of the ground surface and subsurface may generate slope destabilization. From a theoretical perspective, the slope stability depends strongly on the mechanical properties of rocks (Rodriguez-Losada et al., 2009), on the geometry of weakness zones, on the loading and stress conditions (Keefer, 1994; Cala and Flisiak, 2001; Kilburn and Petley, 2003: Verveakis et al., 2007; Urgeles and Camarlinghi, 2013) and on the presence of fluids (Muller-Salzburg, 1987; Crozier et al., 2010; Cappa et al., 2014; Aslan et al., 2021).

In volcanic areas, the interaction between magma reservoirs, magma intrusions, and preexisting faults influences the deformation of volcanic surfaces (Le Corvec and Walter, 2009; Gargani et al., 2006b; Hampel and Hetzel, 2008) and could generate landslides. Phreatomagmatic processes (McMurtry et al., 2004), pore pressure increases due to precipitation (Cervelli et al., 2002) and rift zone intrusions (Le Corvec and Walter, 2009) are believed to influence slope instability on volcanic edifices.

From a practical perspective, present conditions of stability of volcano slopes are influenced by past volcanic activities and geological history; multiple eruptions and landslides, faults and fractures, alteration by fluid circulation, could have generated significant heterogeneities.

Heterogeneities may hide weaknesses and the potential occurrence of slope instabilities. The long geological history of old volcanic edifices is expected to cause complex geometries that are often difficult to understand in detail. On the flanks of volcanoes, the rocks may be constituted by fresh or weathered massive lavas, ignimbrites or low cemented pyroclasts, among other rocks. The past history of the volcanic flanks, as well as the nature of the rocks that constitute their flanks, must be considered carefully when estimating the slope stability of volcanoes from a geotechnical perspective. These complex scenarios could cause unfavorable conditions of stability of the present slope and difficulties in taking into account the spatial heterogeneity of the mechanical parameters.

It has been suggested that giant landslides could be correlated with climatic variations (Mc Murtry et al., 2004). A classical and dramatic example of the role of water is the triggering of the catastrophic Vajont landslide in 1963 in Italy in relation to water-level rise and rainfall increase (Muller-Salzburg, 1987). Some studies assume that giant landslides may have occurred during low stand periods (Quidelleur et al., 2007), whereas others suggest that high stands or sea level rise are more favorable for causing slope instability (Gargani et al., 2014). This question remain controversial and poorly studied despite the expected sea-level rise in relation with climate warming.

The role of climatic conditions on triggering giant landslides on volcanoes is investigated in this study. The possible occurrence of a giant landslide on a volcano in relation to climatic variation is investigated from a theoretical perspective. The processes favoring instability that are discussed in this study are (i) sea-water loading with relative sea-level rise, (ii) pore pressure increase in relation to climatic variation, (iii) rock and soil weathering, (iv) weakness zone geometries, and (v) pressure variation of the magma reservoir caused by sea level unloading.

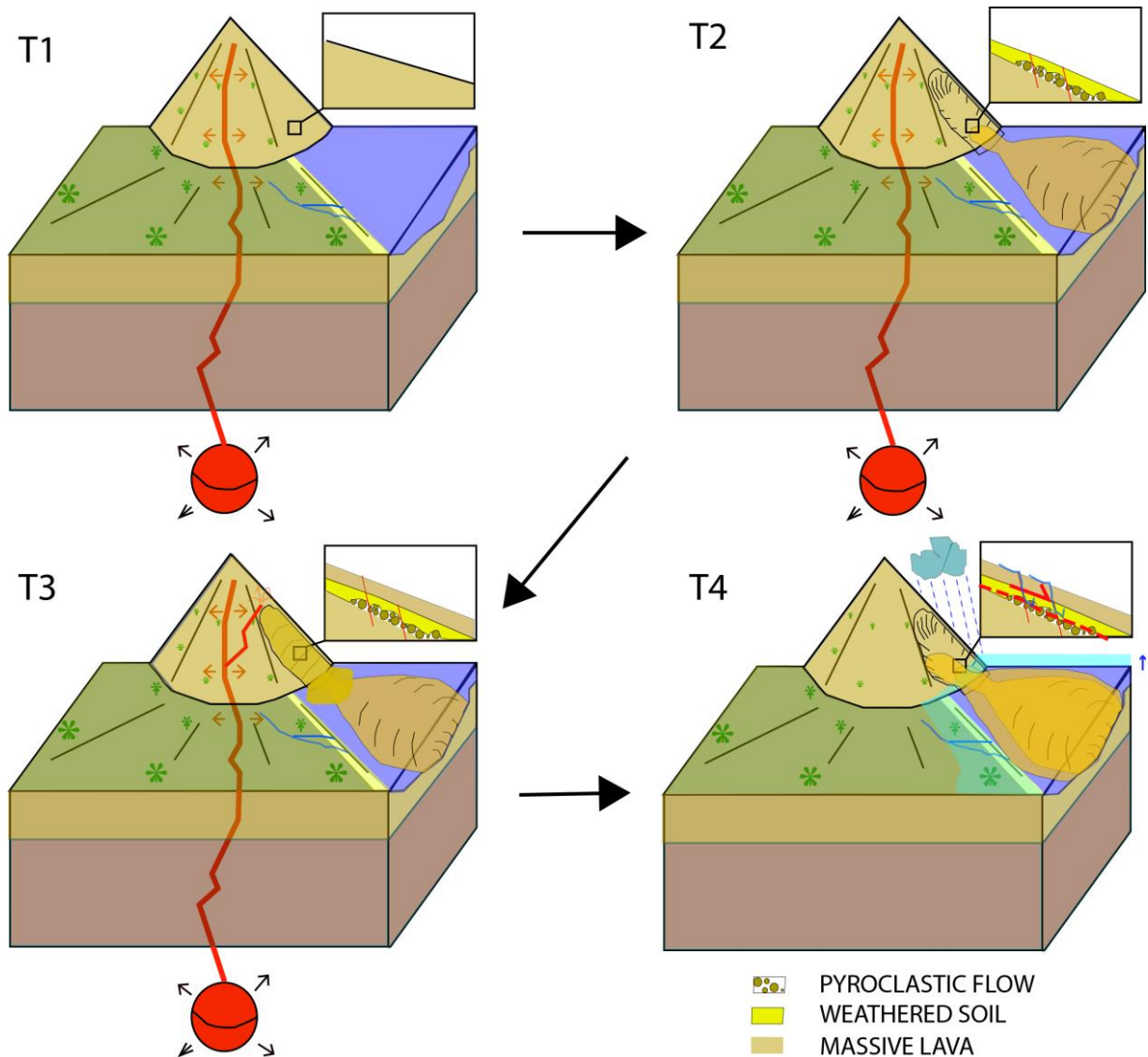

**Figure 1: Complex evolution of volcanic edifices and deep weakness zone locations, (A)** young volcanic edifice composed of massive lavas, **(B)** volcanic edifice after a giant landslide caused by a volcanic process, where pyroclastic flows as well as soil and lava weathering occur, **(C)** filling of the landslide scar by thick volcanic lavas above the pyroclastic debris and the weathered soils, **(D)** deep rooting of the landslide into the weakness zone favored by meteoritic water infiltration or/and sea-level loading in the case of an old and complex volcano.

## 2. Method
### 2.1. Slope stability model

To analyze the causes of stress that may trigger a giant landslide and to discuss the stability of a given relief, a 2D model has been implemented. Two different geometries of the landslide have been employed to compare the role of the geometry of a weakness zone on the theoretical stability of a volcanic edifice. The first geometry (Fig. 2A) is based on the Cullman wedge model (Bigot-Cormier and Montgomery, 2007). Basically, a balance is achieved between the weight of the expected landslide and the force of resistance generated by the rock properties. The resistance shear stress along the slipping surface is given by $\tau_c = \sigma_n tan\phi + C$ until the landslide occurs, where $\sigma_n$ is the normal stress, $\phi$ is the friction angle of the slope-forming material and $C$ is the cohesion. Considering the angle for which the effective cohesion is maximized for a value equal to $\beta/2 + \phi/2$, it is assumed that $\theta = \beta/2 + \phi/2$ (Champel et al., 2002; Bigot-Cornier and Montgomery, 2007). This model allows us to calculate the maximum stable height for a simple geometry (Fig. 2A):

$$H = 4C \cdot \sin\beta \cos\phi / (\rho_r g [1 - \cos(\beta - \phi)])$$

where $H$ is the maximum stable height of the slope, $\beta$ is the hillslope gradient, $\rho_r$ is the bulk density of the rock, and $g$ is the gravitational acceleration. This model has been widely used in geomorphological studies to predict the maximum unfailed height of slopes (Bigot-Cornier and Montgomery, 2007). In the theoretical calculation, vertical loading, pore pressure increase and volcanic rock properties with depth are included. Including the pore pressure U and considering the load of the sea-water column of thickness $H_m$ with a bulk density of the water $\rho_m$, the following can be obtained:

$$H^2 - 4 \cdot H \cdot (C \cdot \sin\beta \cos\phi - U \cdot \sin\phi \cos\beta) / (\rho_r g [1 - \cos(\beta - \phi)]) + \rho_m H_m^2 / \rho_r = 0 \quad (1)$$

Equation (1) allows us to estimate the geometry of the maximum stable relief (height and slope) immediately before the collapse, as well as the geometry of the landslide. The thickness of the sea-water column that is considered could include sea-level variation in relation to Quaternary climatic variation but also relative sea-level increase caused by subsidence.

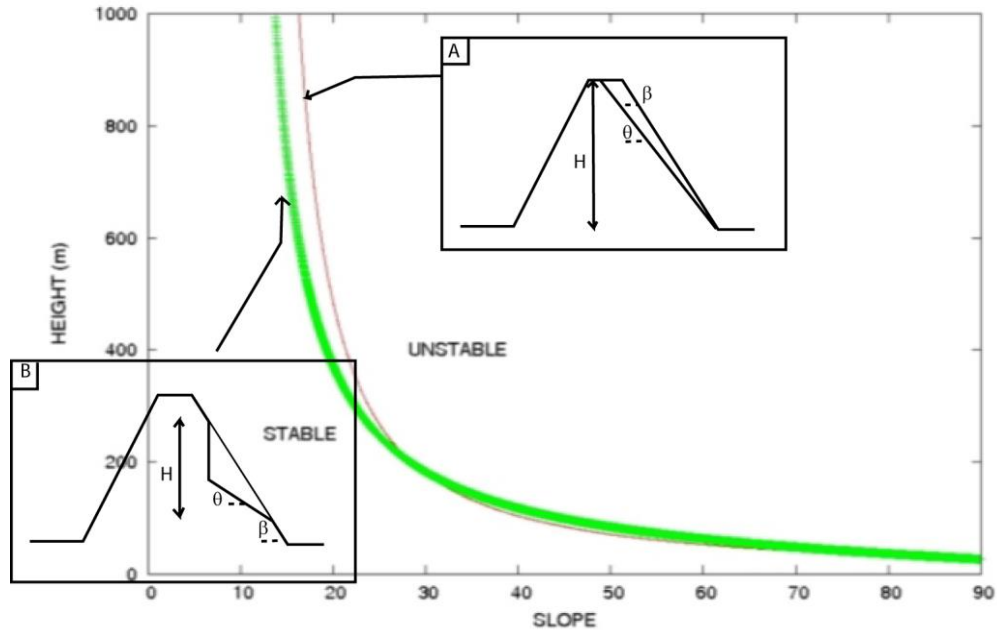

**FIGURE 2: Effect of the initial slope $\beta$ on the maximum stable relief $H$ for two different landslide geometries. (A) Cullman wedge model, (B) concave geometry model. $\rho_c = 3000$ kgm$^{-3}$, $C = 1$ MPa, $\phi = 10^0$.**

Using the same method, it is also possible to calculate the maximum stable height for a second kind of landslide geometry (Fig. 2B, concave geometry model) using the following equation:

$$H = H_1 + H_1 \cos\beta \sin(\beta/2 + \phi/2) / \sin(\beta/2 - \phi/2) \quad (2)$$

where $H_1$ can be calculated by resolving the following equation:

$$H_1^2 - 4H_1 (C \cos\phi - U \sin\phi) / (\rho_r g [\sin\beta - \sin\phi]) + \rho_m H_m^2 \sin(\beta/2 - \phi/2) / (\rho_r \sin\beta \cos\beta/2 + \phi/2) = 0$$

This study focuses on the role of pore pressure, water loading and mechanical properties of rocks on slope stability. Geotechnical approach are also useful to evaluate the safety factor $F_S$ = resisting forces / driving forces ≈ [ $\tau \times S$ ]/[ $\rho_r\ V\ g\ \sin\ \theta$ ] when the slope is destabilized only by his own weight (Hurliman et al., 2000; Cala and Flisiak, 2001; Gargani et al., 2014).

### 2.2. Mechanical properties of volcanic rocks

There are at least two ways to estimate the effective mechanical properties of rocks in a given geological context. The first method is to use experimental data that describe the mechanical behavior of a rock close to that of the studied area. The second is to propose hypotheses on the causes of failure (i.e., gravity, pore pressure, ground acceleration, eruption) and to estimate the effective values of mechanical parameters that triggered the observed failure. Here, we verify that the mechanical properties of rocks estimated using a modeling approach are not in contradiction with experimental data for volcanic rocks.

The parameterization of the model was conducted considering that the relief was close to critical stability. The effective mechanical properties of the rocks of the volcanic edifice considered in this numerical experiment have been calibrated using some characteristics of a giant landslide that occurred 872 kyrs ago in Tahiti (Hildenbrandt et al., 2004 and 2008; Gargani, 2020 and 2022a). However, the investigation presented in this study is purely theoretical, and the geometries as well as the rock heterogeneity do not represent a real case.

The relief is considered to have a total height of ~4500 m with an initial slope $\beta$ of 8-12°. Two different geometries have been tested to estimate the maximum stable relief $H$ as a function of the initial slope (Figure 2). The effect of the precise geometry (Cullman wedge model vs. concave geometry model) of the landslide on the critical value of height and initial slope of the stable relief is not negligible (Figure 2), but this study will not focus on the precise geometry of landslides.

The mechanical properties of volcanic rocks depend strongly on their alteration and stress loading. The cohesion and angle of friction have been estimated for different volcanic rocks under various slope heights by Rodriguez-Losada et al. (2009). The cohesion $C$ can be obtained using the equation $C=kH^l$, where $H$ is the slope height and $k$ and $l$ are coefficients that have been experimentally determined for various volcanic rocks (Rodriguez-Losada et al., 2009). For weathered massive lavas, $k$=0.04 and $l$ varies from 0.41 to 0.60. For low cemented pyroclasts, $k$=0.05 and $l$=0.5. The angle of friction $\phi$ can also be estimated using the equation $\phi=a\ ln(H) + b$, where $a$ and $b$ are obtained experimentally. Rodriguez-Losada et al. (2009) show that very low angles of friction <13° are obtained at depths >1000 m for low cemented pyroclasts ($a$=6; $b$=54). A volcanic edifice constituted by low cemented pyroclasts has a cohesion between 0.5 and 1 MPa at depths of 1500-3000 m (Figure 3a). At this depth, a theoretic volcanic edifice constituted by low cemented pyroclasts has an angle of friction that ranges between 6° and 10° (Figure 3b).

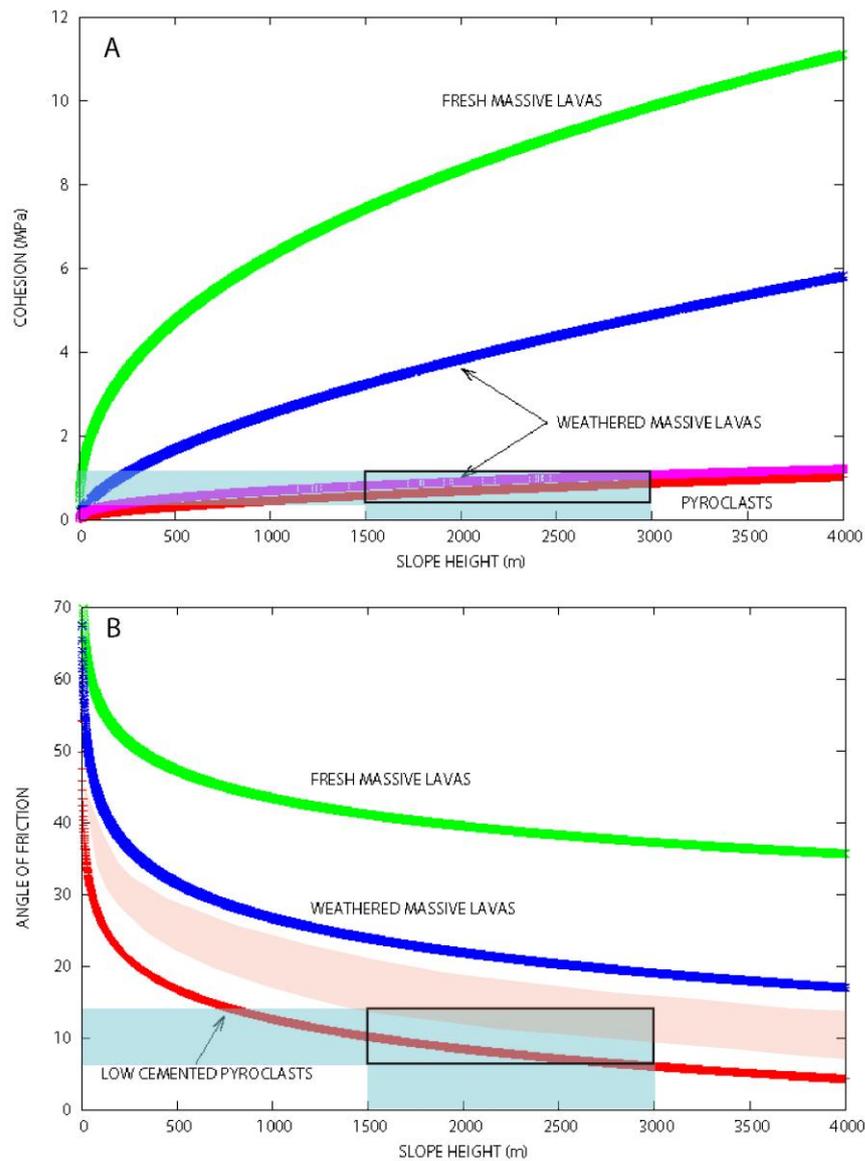

**FIGURE 3: Effect of depth on mechanical properties. (A)** Cohesion for low cemented pyroclasts ($k$=0.05; $l$=0.5) and fresh ($k$=0.37; $l$=0.60) and weathered ($k$=0.04; $l$=0.41 and 0.60) massive lavas, **(B)** angle of friction for low cemented pyroclasts ($a$=6; $b$=54) and fresh ($a$=5.6; $b$=82) and weathered ($a$=7; $b$=75) massive lavas.

### 2.3. Modeling hydroclimatic variations

In this study, sea-level loading and pore pressure variation in relation to climatic evolution were simulated. Basically, the water column loading that is considered has been estimated using a proxy. Sea level is known to change contemporaneously with $\delta^{18}O$ variation (Gargani and Rigollet, 2007; Gargani et al., 2008). To simulate climate forcing, a $\delta^{18}O$ curve (Lisiecki and Raymo, 2005) was used. This curve has been recalibrated to simulate Quaternary sea-level variations that reach an amplitude of 120 m.

The hydroclimatic variations are also caused by Milankovitch astronomic cycles and can be simulated using insolation or $\delta^{18}O$ curves (Gargani et al., 2006a). In this study, it is considered that more humid conditions are able to cause an increase in the effective pore pressure. The aim of this assumption is to discuss the timing of potential giant landslides in relation to Quaternary climatic conditions (i.e., precipitation rates). The time necessary for the water to propagate from the surface to a depth of 4.5 km has been estimated at ~150 days in Oregon (Saar and Manga, 2003). In the model, the propagation is considered instantaneous, which is justified on a long time scale (>1 kyr). The amplitude of the variation in the pore pressure ΔP due to water infiltration into the crust is 0.01 MPa at Mt. Hood (Saar and Manga, 2003) and 2 MPa on the south flank of Kilauea volcano (Cervelli et al., 2002). In this study, various values for the pore pressure variation have been tested in the case where sea-level effects are dominant (0 < ΔP < 0.125 MPa) and in the case where pore pressure processes are dominant (0 < ΔP < 0.5 MPa).

### 2.4. Pressure variation in the magma reservoir

The loading variation caused by sea level variation is $\Delta L_{SL}(t) = g\, \rho_m\, H_m(t)$.

Sea water unloading causing decreased lithostatic pressure at depth, enhances the production of magma. Decompression favor partial mantle melting and magma release. The magma production rate $DF/Dt$ at constant entropy $S$ can be estimated as (Jull and McKenzie, 1996; Sternai et al., 2017):

$DF/Dt = (\delta F/\delta P_M)_S\, (dP_M/dt - U \cdot \nabla P_M)$

where $P_M$ the magma reservoir pressure, $F$ the melt fraction, $t$ the time, $U$ the mean mantle upwelling rate.

The magma pressure variation $\delta \Delta P_M/dt$ can be considered as depending from the sea water unloading variation $\delta \Delta L_{SL}/dt$ under adiabatic condition and for very high viscosity (i.e. $\eta_c = 10^{23}$ Pa.s) of the crustal rocks. In this case $\delta \Delta P_M/dt = - \delta \Delta L_{SL}/dt$.

When the viscous response is not negligible ($\eta_c < 10^{22}$ Pa.s; Sternai et al., 2017), the equation is $\delta \Delta P_M/dt + \Delta P_M (E_c/\eta_c) = - \delta \Delta L_{SL}/dt$, where $E_c$ is the elastic modulus, and a delay of ~10 kyr after the forcing by sea water unloading is expected.

### 3. Results

The current question that we attempt to answer focuses on the thresholds and conditions that favor slope instability for volcanic edifices in relation with climatic variations.

### *3.1. Influence of water column loading on slope stability*

Considering a simple geometry for the expected landslide (Fig. 2A, Cullman wedge geometry), the role of sea level variation on slope stability was investigated using equation (1). When a volcanic island subsides, an increase in the relative sea level and, consequently, an increase in the water loading on the slope occur and could be significant.

Under the Cullman wedge geometry condition, an increase in the loading by a water column of 120 m cause a slight decrease in the stability height (Figure 4C). The more the water loading increases, the less the stability. Whatever the geometry of the sliding mass is, the maximum height of the stable relief is impacted by the increase in the loading by a water column of 400 m. A water column increase of 400 m is reached in less than 1 Myr for a moderate subsidence rate (~0.25 mm/yr) contemporaneously with sea-level high stand at +120 m during an interglacial period. In this case, slope instabilities can be observed on higher slopes (>25°) but also on smaller slopes (Figure 4). The role of sea-water loading could trigger landslides of smaller dimensions (150-300 m) for higher slopes (25-30°, Fig. 4C, dark gray square).

## 3.2. Influence of the pore pressure on slope stability

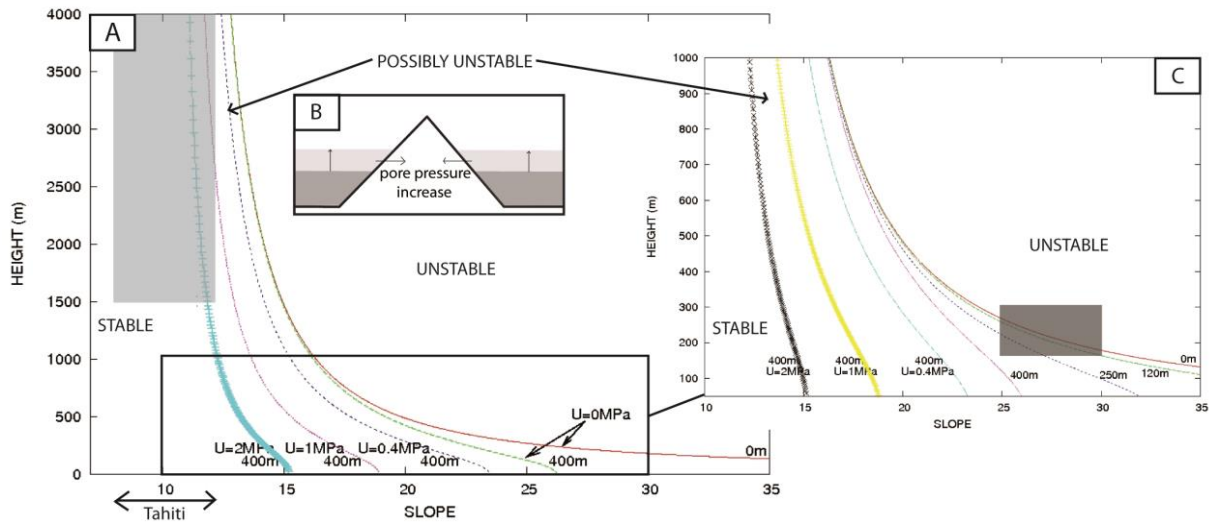

**FIGURE 4:** (A) Effect of the increase in pore pressure and sea-water loading on the stability of the relief. The light gray area is the area that changes from a theoretically stable relief to an unstable relief when submitted to an increase in the sea-water column loading between 0 and 400 m as well as to a pore-water pressure increase between 0 MPa and 2 MPa. (B) Schematic representation of the effect of subsidence. (C) The details are as follows: a relief of ~200 m with a slope of $25° < \beta < 30°$ could be affected by water loading (dark gray square). $\rho_c$ = 3000 kgm$^{-3}$, $C$ = 1 MPa, $\phi$ = 10° and geometry of the landslide identical to those presented in Figure 2A.

The model suggests that a significant effect of the pore pressure increase on the height of the stable relief could occur (Fig. 4). The pore pressure significantly reduces the stability of the relief. As expected, the higher the pore pressure is, the lower the maximum height $H$ before the giant landslide occurs. If the pore pressure increases by more than 1 MPa, the slope stability is significantly decreased (Fig. 4A and B). A relief of more than 1500 m, with a slope of ~12°, could be affected by landslides under these conditions. Depending on the initial relief $H$, a pore pressure increase of 1 MPa could cause instabilities of more than 1000 m.

### 3.3. Competing influence of sea-level loading vs. pore pressure variations over time on slope stability

In this study, the influence of sea-level loading and pore pressure variations over time, in relation to meteoritic water infiltration, was modeled. The variation over time of the maximum height $H$ of the volcanic edifice before any giant landslide occurs is shown in Figure 5. The maximum stable height depends on the loading of the sea level and on the value of the pore pressure. Taking into account that the maximum sea-level variation during the last million years is ~120 m and that the amplitude and the timing of the loading are known (Figure 5B), the variation in the maximum stable height $H$ of a theoretical volcano before any giant landslide occurs has been calculated. The absolute value of the maximum stable height $H$ strongly depends on mechanical properties of rocks and slope, and will not be discussed in this study because this study focus on the variation of stability caused by climatic evolutions.

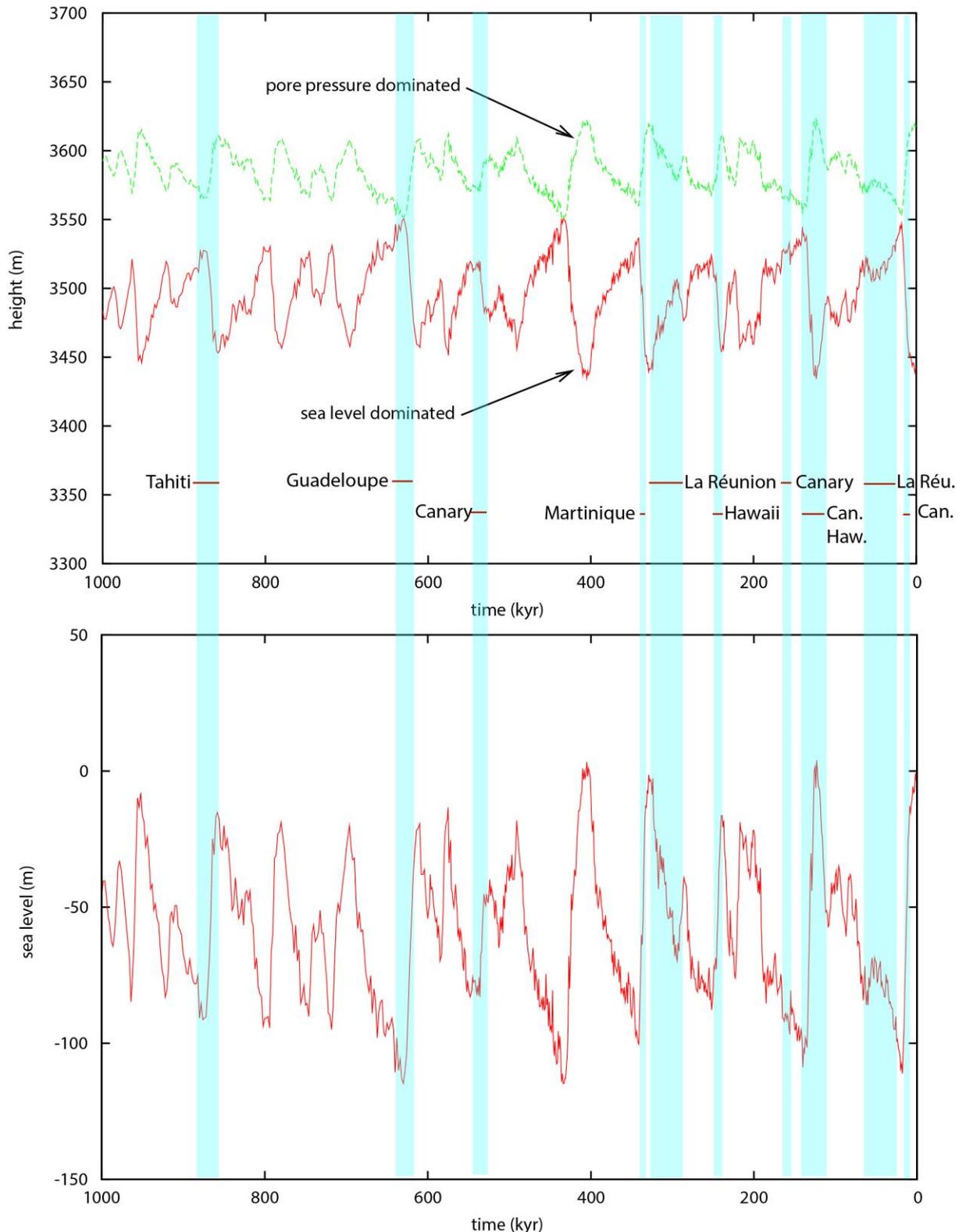

**FIGURE 5:** (A) Maximum height H before the landslide occurred. The geometry of the island is shown in Figure 4. The effects of sea-level loading and pore pressure are simulated. The sea-water loading depends on the variation in sea level, whereas the effect of the pore pressure depends on the variations in precipitation. In the first case (sea-level-dominated), the pore pressure is considered to range between 0 and 125 kPa depending on the climatic variation. In the second case (pore pressure-

**dominated), the pore pressure is considered to range between 0 and 500 kPa depending on the climatic variation. The absolute values of H is indicative and only the trends are interpreted in this study. $g$=9.81 ms$^{-2}$, $C$=1 MPa, $\phi$=10°, $\rho$=2800 kg/m$^{-3}$, $\beta$=7°, and a subsidence rate of 0.25 mm/yr are considered. (B) Simulated sea-level variations during the last million years using the $\delta^{18}O$ curve of Lisiecki and Raymo (2005). The light blue lines represent the giant landslides on volcanoes during the last 1 Ma of Table 1.**

When seawater loading is the main process that influences landslides, the higher the sea level is, the lower the stability. The maximum height $H$ is expected to decrease when the sea level rises (Figure 5). Considering all other parameters as identical, the maximum stable height $H$ of a volcanic edifice is 50 m higher during interglacial periods than during glacial periods if the sea-water loading is the main parameter influencing the system.

In the specific case of Polynesian climate variation, where the climate is more humid during glacial maxima (Saéz et al., 2009), as simulated here, the giant landslides generated by a pore pressure increase are expected to occur during glacial periods (Figure 5A). In the specific case of the Polynesian volcanic edifice, the period when the landslides are expected (i.e., the time when the maximum stable height $H$ before a giant landslide is expected to occur), is different when the dominant process that triggers the landslide is sea-level loading or pore pressure loading (Figure 5A).

The influence of sea level unloading on magma reservoir pressure increases at depth is better correlated than the sea level loading with giant landslides occurrence (Fig. 5). This could suggest that in many cases the volcanoes were actives (*i.e.* a magma reservoir was present at depth). If this process was the cause of the giant landslides, the viscosity of the crust was very high in many cases, except in La Réunion (290-320 kyr; Leunat and Labazuy, 2008), Canary islands (134 ± 6; Carracedo et al., 1999) and Hawaii (127 ± 10; McMurtry et al., 1999) where a delay is observed with low sea level. The effect of sea level unloading influence on magma pressure increase is contemporaneous with the effect caused by pore pressure increases with meteoritic water infiltration in Polynesia and East Africa (*i.e.* where precipitation increases during glacial periods).

## 4. Discussion
### 4.1. Rock weathering and slope instability

Volcanic edifices are usually composed of massive lavas at the surface. These lavas can be highly weathered by the humid climate when they are exposed for thousands or millions of years. Laboratory experiments for weathered massive lavas show that cohesion can be low (0.8-1 MPa) at depths ranging from 1500 m to 3000 m (Figure 3a).

The angle of friction of weathered massive lavas is estimated to range between 23° and 19° based on laboratory experiments (Figure 3b). Lavas that have been extremely altered by rain and fluid circulation at depth for thousands of years are expected to have mechanical properties lower than those of fresh volcanic rocks. Furthermore, the effective mechanical properties ($C$ and $\phi$) of fractured and faulted reliefs are lower than those used for triaxial experiments on nonfractured rocks. Effective mechanical properties take into account the heterogeneities present in old and complex volcanic edifices including low cemented pyroclast and weathered soils. Soils can be observed sandwiched between volcanic lavas (Hevia-Cruz et al., 2022). These soils are expected to have reduced mechanical properties. The model has been calibrated to be compatible with past giant landslide events (~1500-3000 m thick, initial slope $\beta$ ~8-12°, cohesion $C$ of 0.3-1 MPa, Figure 2), such as the event that occurred in Tahiti, Society Archipelago (Gargani, 2020 and 2022a) (Figure 2). The values used here correspond to highly weathered rocks of an old and fractured volcanic edifice or low cemented pyroclasts that could be sandwiched at depth.

Rock mechanical properties $C$ and $\phi$ decrease over time when exposed to weathering. The more weathering that occurs, the greater the decrease in $C$ and $\phi$. During humid periods, this decrease is higher than during drier periods (Figure 6A). In contrast, diagenetic processes are expected to improve the mechanical properties of rocks (Figure 6B). When pressure increases at depth, porosity and connectivity can decrease. Furthermore, mineralization of fluid at depth can also improve rock cohesion in specific cases.

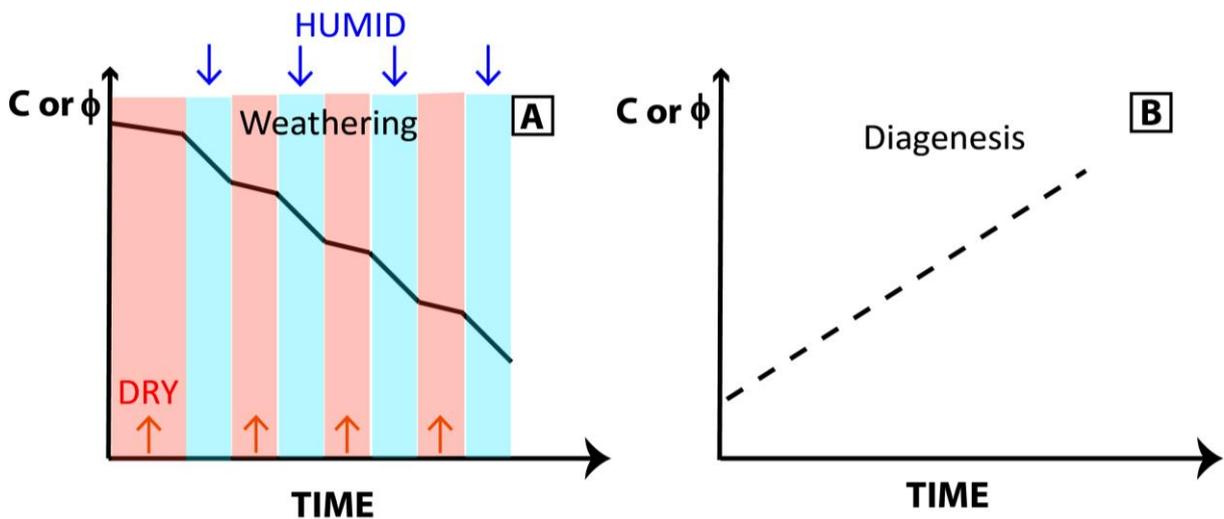

**FIGURE 6: Influence on the rock mechanical properties C and $\phi$ of (A) weathering and (B) diagenesis.**

*4.2 Weakness zones and geological inheritance*

The construction of volcanoes is often polygenic, and numerous heterogeneities exist on volcanic edifices (Hildenbrand et al. 2006). The complex and long volcano-tectonic history of old volcanic edifices may explain why numerous fractures and significant weathering are observed. The highly fractured and heterogeneous material (basaltic flows and volcanic breccia) of volcanic edifices is often complex and difficult to model in detail at all scales. Strata of weathered soils sandwiched between lavas can also be observed (Hevia-Cruz et al., 2022) as well as pyroclastic formations. Low values for the angle of friction have been estimated or used in various studies for weakness structures, effective rock properties or faults (Cala and Flisiak, 2001; Bigot-Cormier and Montgomery, 2007; Got et al., 2008; Egholm et al., 2008; Abers, 2009).

Modeling studies are used to simplify the complexity of the real world. The effective mechanical properties implemented in modeling are different than the laboratory experimental properties. Preexisting zones of weakness at depth are present in many volcanic edifices but are different in each case. A real geometry of a specific volcanic edifice or of a specific landslide is not the aim of this study. A general case is considered. This model does not permit us to predict exactly when a large landslide is expected to occur but only to discuss the possibility of discriminating the influence of meteoritic water infiltration from sea-water loading/unloading using climatic correlation from a theoretical perspective.

*4.3 Sea-level variation effect on landslides and volcanic activity*

The role of water-level rise in landslide triggering has been suggested in the case of the Vajont landslide (1963, Italy) (Muller-Salzburg, 1987; Kilburn and Petley, 2003). When sea level rises, (i) water loading increases and (ii) water infiltrates into rocks. The loading caused by sea level rise could cause slope instability when the slope is almost at equilibrium immediately before sea level rise. Furthermore, when sea level rise, erosion can occur at the base of the cliff and favor cliff retreat, rock fall and landslide (Ye et al., 2013). However, the effect of loading forces on the base of a slope can also favor the slope stability. A vertical loading force at the base of the volcano can cause an increase of stability by opposing a force to the potential movements of the relief. In other words, when the sea-level fall at the base of a slope, it could favor instability (Figure 7A). This result is suggested by safety factor $F_S$ decrease when loading decreases. When a significant loading is above the center of mass of the potential landslide, it cause the destabilization of the slope by triggering rock failure (Figure 7B).

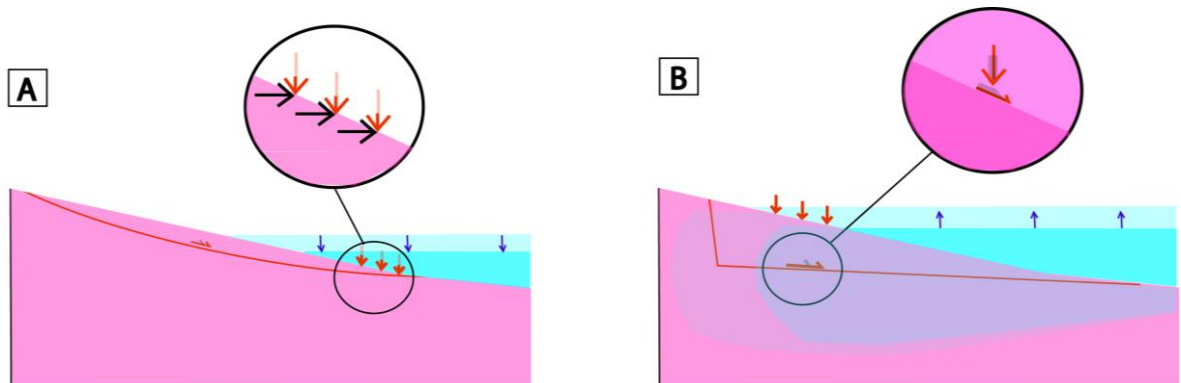

**Figure 7: Influence of the sea-level variation in relation with the position of the center of mass of the potential sliding area: (A) slope instability caused by sea-level lowering when the load is located at the base of the slope, (B) slope instability caused by sea-level rise above the center of mass with marine water infiltration.**

The giant landslides in volcanic areas seem to be correlated with climatic variation (Quidelleur et al., 2008). Sea level loading is one of the potential causes of giant landslides (Aslan et al., 2021). However, indirect interactions are also possible in volcanic areas. The unloading associated with sea level lowering could modify the pressure on the magma reservoir (Sternai et al., 2017). A pressure decrease has implications for the input and output rates of magma into and out of the magma reservoir. If the connection to the deeper magma source remains open, one consequence is an increased rate of replenishment with primitive magma (Pinel and Jaupart, 2003). Furthermore, a reduced load allows the eruption of denser magmas that would otherwise have been stuck at shallow depths (Pinel and Jaupart, 2000).

The potential interaction between sea-level variation and landslides on one side and landslides and volcanic eruptions on the other side (Longpré et al., 2009) could cause also difficulties in interpretation. If the magma reservoir is shallow, a landslide could generate dyke intrusion, but if the magma reservoir is deep, dyke intrusion is more difficult after a landslide.

*4.4 Pore pressure variation*
Another process that is able to cause landsliding is pore pressure increase. Fluids are known to be a triggering and driving factor for landslides (Cappa et al., 2014). Various mechanisms could explain a pore pressure increase of ~1 MPa. First, it is necessary for a significant amount of water to enter the system. For example, unsaturated rocks of porosity $n$ in an area of volume $V$ could become saturated when a significant water volume $n \cdot V$ infiltrates. The volume of water could be higher if a dense fracture network existed. Meteoritic water infiltration at a depth of 1000 m under lithostatic pressure could cause a pore pressure increase. An increase of more than 1 MPa in the pore pressure at depth related to meteoritic water is not unrealistic and has been proposed in previous studies (Cervelli et al., 2002).

An alternative mechanism that could also lead to a significant pore pressure increase results from the progressive collapse of the base of the volcanic edifice. In this case, the slow creep of the edifice on a preexisting weak structure may trigger overpressure conditions locally that reach high pressures (Veveakis et al., 2007). In a volcanic context, it may also be expected that hydrothermal processes play a role in the pore pressure increase and rock alteration at depth.

The pore pressure can change in relation to meteoritic water infiltration in highly fractured rocks. In this case, meteoritic water infiltration increases when the precipitation rate increases. Consequently, climatic variations are expected to influence the pore pressure variation in highly fractured rocks. The correlation between climatic variation and giant landslides may be caused by pore pressure increases at depth.

*4.5 Climate variation and correlation with giant landslides*
Precipitation increased during glacial maxima in Polynesia (Saez et al., 2009), Mexico (Ganeshram and Pedersen, 1998), and eastern equatorial Africa (Chiang, 2009) but decreased in NW Europe (Guiot et al., 1989), the Caribbean (Curtis et al., 2001) and Indonesia (Costa et al., 2015; Russell et al., 2014).

When humid conditions occurred during warmer phases, the potential effect of water infiltration on pore pressure acted to increase instability during warmer interglacial periods, similar to sea-level loading. Interglacial phases are expected to cause more slope instability in this case. However, it is difficult to discriminate the influence of sea-level loading from meteoritic water infiltration at depth on giant paleolandslides in this case.

In contrast, when precipitation increased during glacial maxima, the impacts of sea-level loading and water infiltration at depth on giant paleolandslides were not contemporaneous. In this case, it is theoretically possible to discriminate the process that plays the main role in triggering giant landslides. For example, in Polynesia, increased meteoritic water infiltration occurred during glacial maxima and potentially increased slope instability during glacial maxima, whereas sea-level rise caused slope instability during interglacial times (Figure 5A). The age of giant landslides (Table 1) may be a good argument to discriminate between these two processes, but only in Polynesia, Mexico or eastern equatorial Africa and not in NW Europe, the Caribbean or Indonesia.

This study suggests that if preexisting weakness zones are present at depth, causing low effective cohesion and internal friction, a climatic origin of landslides is possible even if volcano is extinct. Seismic or volcanic processes are not necessarily directly responsible for giant landslides, even if these processes may have played a role in causing weakness zones in the past in these areas. In 70% of cases, the giant landslides occurred when sea-level rise, regardless the precipitation rate is. Due to potential delay caused by viscous processes (i.e. 10 kyr of delay for a viscosity less than $10^{22}$ Pa.s), it could be also correlated to the magma chamber unloading by sea-level lowering. This result could be influenced by the set of data considered and further studies on giant landslides on volcanic islands would improve this interpretation.

**Table 1**: *Giant landslide ages in volcanic areas during the last 1 Ma.*

| Volcano | Age (ka) | References |
|---|---|---|
| Canary Islands (El Hierro, El Golfo) | 10-17 | Gee et al., 2001 |
| La Réunion | 20-68 | Lenat and Labazuy, 2008 |
| Hawaii (Alika phase 1) | 112 | Mc Murtry et al., 1999 |
| Hawaii (Alika phase 2) | 127 $\pm$ 10 | Mc Murtry et al., 1999 |
| Canary Islands (El Hierro, El Golfo) | 134 $\pm$ 6 | Carracedo et al., 1999 |
| Hawaii (Southern Lanai) | 135 | Rubin et al., 2000 |
| Canary Islands (Tenerife) | 150-170 | Hunt et al., 2011; Masson et al., 2002 |
| Hawaii (Southern Lanai) | 240 | Rubin et al., 2000 |
| La Réunion | 290-320 | Lenat and Labazuy, 2008 |
| Martinique | 337 $\pm$ 5 | Quidelleur et al., 2004 |
| Canary Islands (La Palma) | 537 $\pm$ 8 | Guillou et al., 2001; Groom et al., 2022 |
| Guadeloupe | 629 $\pm$ 13 | Samper et al., 2007 |
| Hawaii (Haleakala, Hana) | 860 | Moore and Clague, 1992 |
| Tahiti-Nui (north) | 872 $\pm$ 10 | Hildenbrandt et al., 2004 |

### 4.6 Small landslides vs. giant landslides

The dimension of the expected landslide depends on the initial slope of the relief. Locally, as in a deeply incised canyon of volcanic islands, significant slopes can be observed. In the case of a significant slope (>25°), the critical value for the height of the stable relief is less than 300 m (Fig. 3C). Consequently, smaller landslides may also be triggered locally. When rotational landslides occur, significant slopes develop in the upstream part. The resulting relief could be near the instability conditions that favor the development of new landslides and other retrogressive erosion processes. It can be difficult to discriminate the relief associated with numerous "small" landslides from that caused by giant landslides (Gargani, 2020; Gargani, 2022b).

Erosion and repetition of small landslides could progressively reduce the loading. As previously shown, the loading influences the maximum height expected before a landslide occurs. Reducing the load increases the maximum height before landsliding and thus reduces the potential occurrence of a landslide when volcanic edifice is an extinct volcano. Consequently, the potential occurrence of a giant landslide could be reduced by the occurrence of small landslides and erosion of the relief in the case of extinct volcanoes. Previous studies have shown that the probability of small landslides occurring is higher than that of large landslides (Urgeles and Camerlinghi, 2013). Nevertheless, giant landslides have also been described, suggesting that there are specific conditions that favor giant landslides.

## 5. Conclusion

The probability of a large landslide in an area where no significant volcanic and seismic activities have been observed during the last thousand years is not null. Large landslides caused by nonvolcanic processes could occur on old volcanic islands. Indeed, our modeling suggests that a large landslide may occur due to an increase in pore pressure and/or sea-level variation. These processes could be discriminated in areas where glacial maxima are more humid than interglacial periods, but not in the other cases, based only on timing correlation. This finding is also a consequence of the mechanical properties of highly weathered and/or fractured volcanic rocks in weakness zones (low cohesion and angle of friction) at depth in old weathered volcanic edifices. When volcanoes are still actives, sea level unloading can be responsible of giant landslides, causing magma reservoir pressure increases at depth and magma release. During the last million years, several giant landslides in tropical areas are correlated with sea level unloading during glacial periods rise suggesting that this effect is a driving mechanism.

**Acknowledgments:**